\input{aipcheck}

\documentclass[
  ,draft            
  ]
  {aipproc}

\layoutstyle{8x11double}

\newcommand {\snn}      {\sqrt{s_{\rm NN}}}
\newcommand {\raa}      {R_{\rm AA}}

\newcommand {\pt}       {p_{\rm T}}

\begin{document}

\title{Heavy-flavour production in hot quark matter}



\author{Andr\'e Mischke\footnote{Email: a.mischke@uu.nl}}{
  address={ERC-Starting Independent Research Group QGP - ALICE, Institute for Subatomic Physics,
Utrecht University, Princetonplein 5, 3584 CC Utrecht, the Netherlands.}}

\begin{abstract}
Recent results from the RHIC facility and first measurements from the ALICE experiment at the CERN-Large Hadron Collider (LHC) on open heavy-flavour are presented. 
We focus on RHIC measurements of single electrons and jet-like heavy-flavour particle correlations. 
First D meson signals from 7 TeV proton-proton collisions from ALICE are discussed.
Next-to-leading-order QCD processes, such as gluon splitting, become important at LHC energies and its contribution can be accesses
by the measurement of the {\em charm content in jets}.
\end{abstract}

\maketitle

\section{Introduction}
High-energy nucleus-nucleus collisions allow exploring the behaviour of QCD matter at high temperatures, where a new phase of matter, the Quark-Gluon Plasma (QGP), is predicted to exist. Measurements at the RHIC facility suggest that a dense,
equilibrated system of quarks and gluons is created in such collision with properties similar to that of an ideal hydrodynamic fluid. The strong suppression of hadron production at high transverse momentum indicates that the system early in its evolution is extremely dense and dissipative~\cite{pbm07}.

Especially, heavy quarks provide sensitive penetrating probes of the medium. Due to their large mass, heavy quarks are believed to be predominantly produced in the initial state of the collision by gluon fusion processes so that they probe the entire lifetime of the QGP. Theoretical models have predicted that heavy quarks should experience smaller energy loss than light quarks while propagating through the QCD medium due to the suppression of small angle gluon radiation, the so-called 
{\em dead-cone effect}~\cite{deadcone1, deadcone2}.
The study of heavy-flavour production in nucleus-nucleus collisions provides key tests of parton energy-loss models and, thus, yields profound insight into the properties of the produced matter~\cite{tho08}.

\section{Results from RHIC}
Both STAR and PHENIX have measured open heavy-flavour production in different ways. Besides the direct reconstruction of D mesons~\cite{tho08}, which is currently restricted to low $\pt$, 
charm and bottom quarks are identified by assuming that isolated electrons in the event stem from semi-leptonic decays of heavy-quark mesons. At high $\pt$ this mechanism of electron production is dominant enough to reliably subtract other sources of electrons like conversions from photons and $\pi^0$ Dalitz decays.

\begin{figure}[t]
\centering
   \includegraphics[scale=0.3]{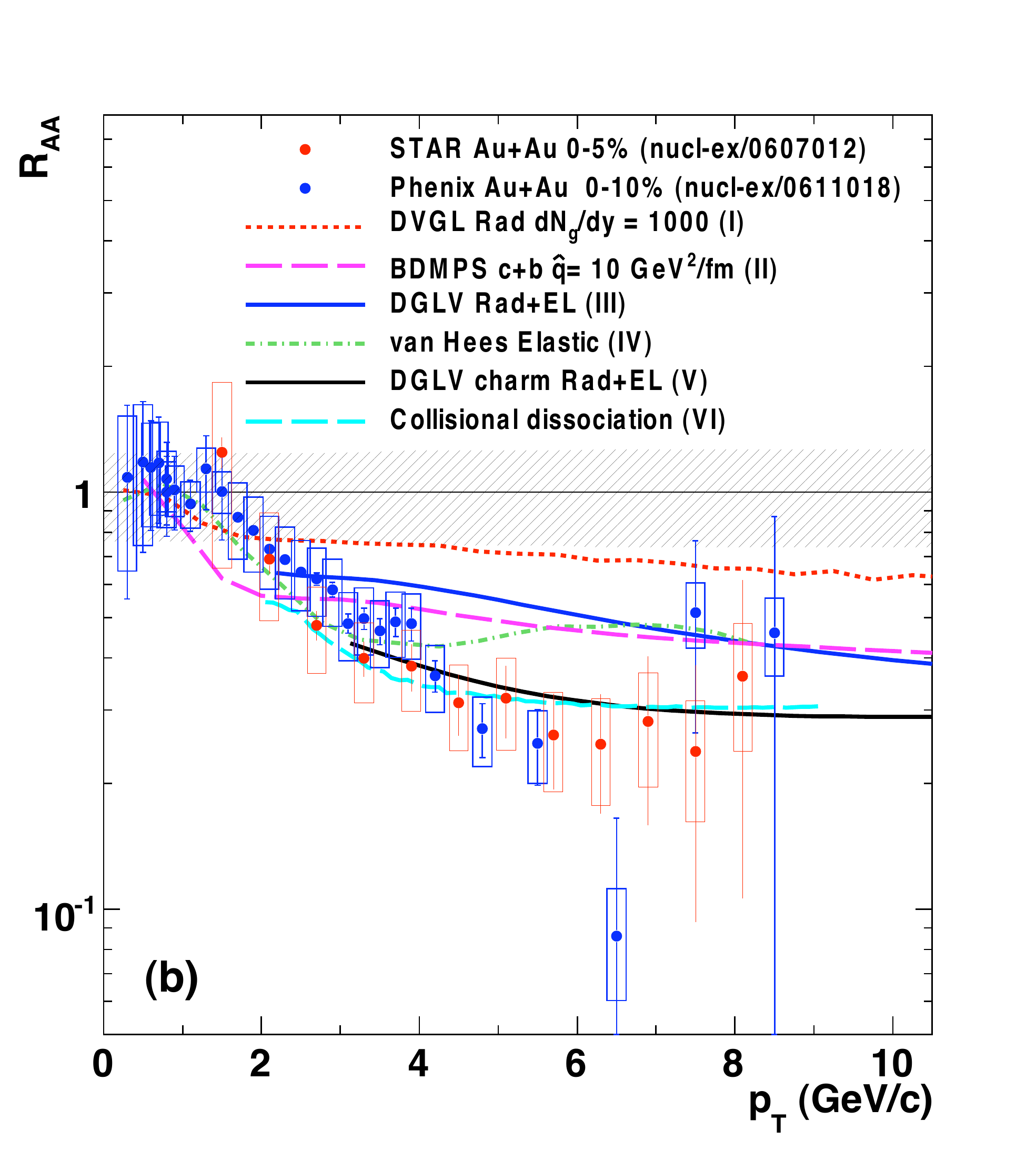}
   \includegraphics[scale=0.4]{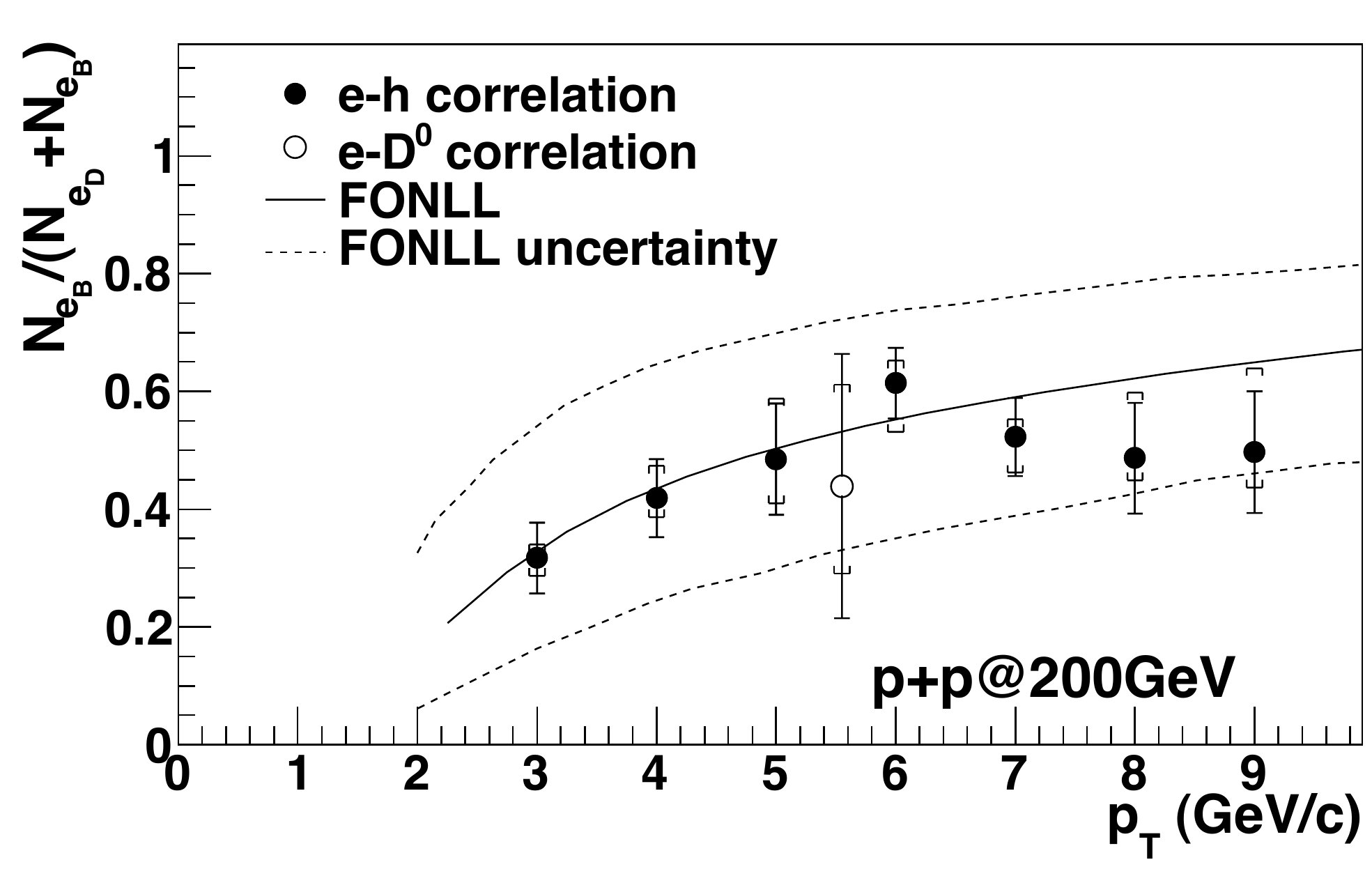}
   \caption{Left: Nuclear modification factor $\raa$ of heavy-flavour decay electrons in most central Au+Au collisions at $\snn$~= 200 GeV. The curves indicate different theoretical model calculations. Right: Relative bottom contribution to the total single electron yield obtained from e$-D^{0}$ and e$-$hadron correlations, compared to the uncertainty band from pQCD calculations at Fixed-Order plus Next-to-Leading Logarithm (FONLL) level~\cite{Mat05}.}  \label{Fig:1}
\end{figure}

Nuclear effects are typically quantified using the nuclear modification factor $\raa$ where the particle yield in Au+Au collisions is divided by the yield in $p$+$p$ reactions scaled by the number of binary collisions. 
$\raa$~= 1 would indicate that no nuclear effects, such as Cronin effect, shadowing or gluon saturation, are present and that nucleus-nucleus collisions can be considered as an incoherent superposition of nucleon-nucleon interactions.
Figure~\ref{Fig:1}, left panel, shows the $\pt$ dependence of $\raa$ for single electrons in central Au+Au collisions from 
STAR~\cite{Star:npe} and PHENIX~\cite{Phe:npe}, which are consistent with each other. 
Surprisingly, the suppression of the single electron yield is as strong as observed for light quark hadrons (factor of $\sim$5), in contradiction to expections from the dead-cone effect~\cite{deadcone1, deadcone2}.
The observed suppression is overpredicted by theoretical model calculations using reasonalbe input parameters.
The data are described reasonably well if the bottom contribution to the electrons is assumed to be small. 
Therefore, the observed discrepancy could indicate that the $B$ dominance over $D$ mesons starts at higher $\pt$.
A possible scenario for $B$ meson suppression invokes collisional dissociation in the medium~\cite{Theo:Adil}.

However, it has been found~\cite{Phe:ehpaper, Star:ehpaper} that bottom contributes significantly ($\sim$50$\%$) to the single electron yields above $\pt$~= 5~GeV/$c$ (cf. Fig.~\ref{Fig:1}, right panel). 
More systematic measurements will be possible with the installation of high precision vertex detectors in STAR and PHENIX, which will allow the direct measurement of open charm and bottom mesons.
\begin{figure}[th!]
\centering
   \includegraphics[scale=0.58]{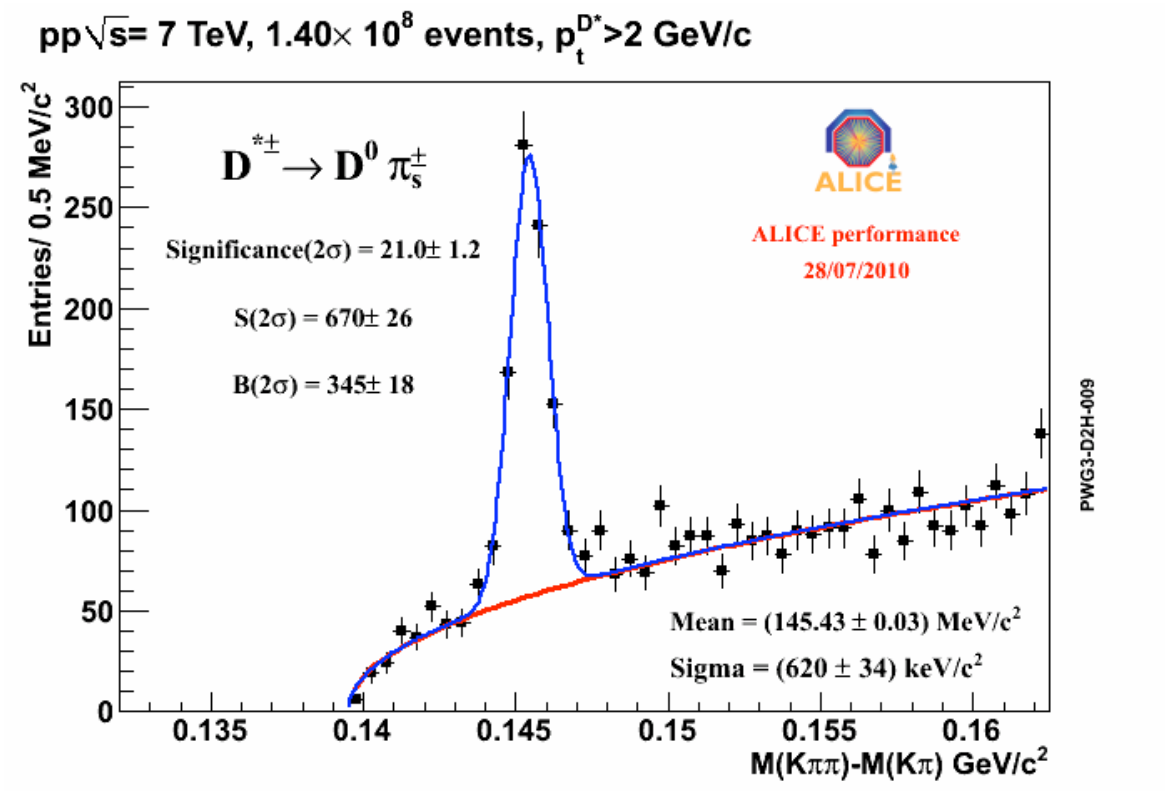}
   \includegraphics[scale=0.6]{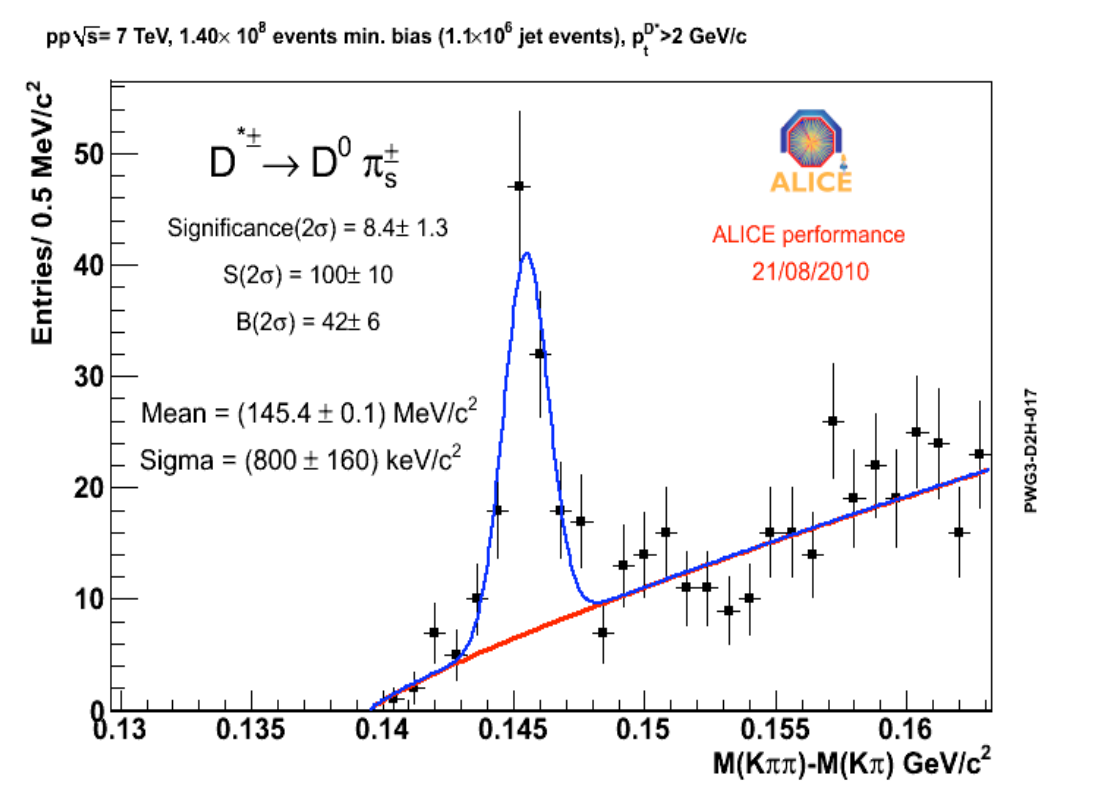}
   \caption{$D^*$ signal in 7 TeV minimum bias $p+p$ collisions (left panel) and reconstructed within the cone of jets (right panel). See text for details.}  \label{Fig:4}
\end{figure}

\section{First results from ALICE}
\begin{figure}[t]
\centering
   \includegraphics[scale=0.5]{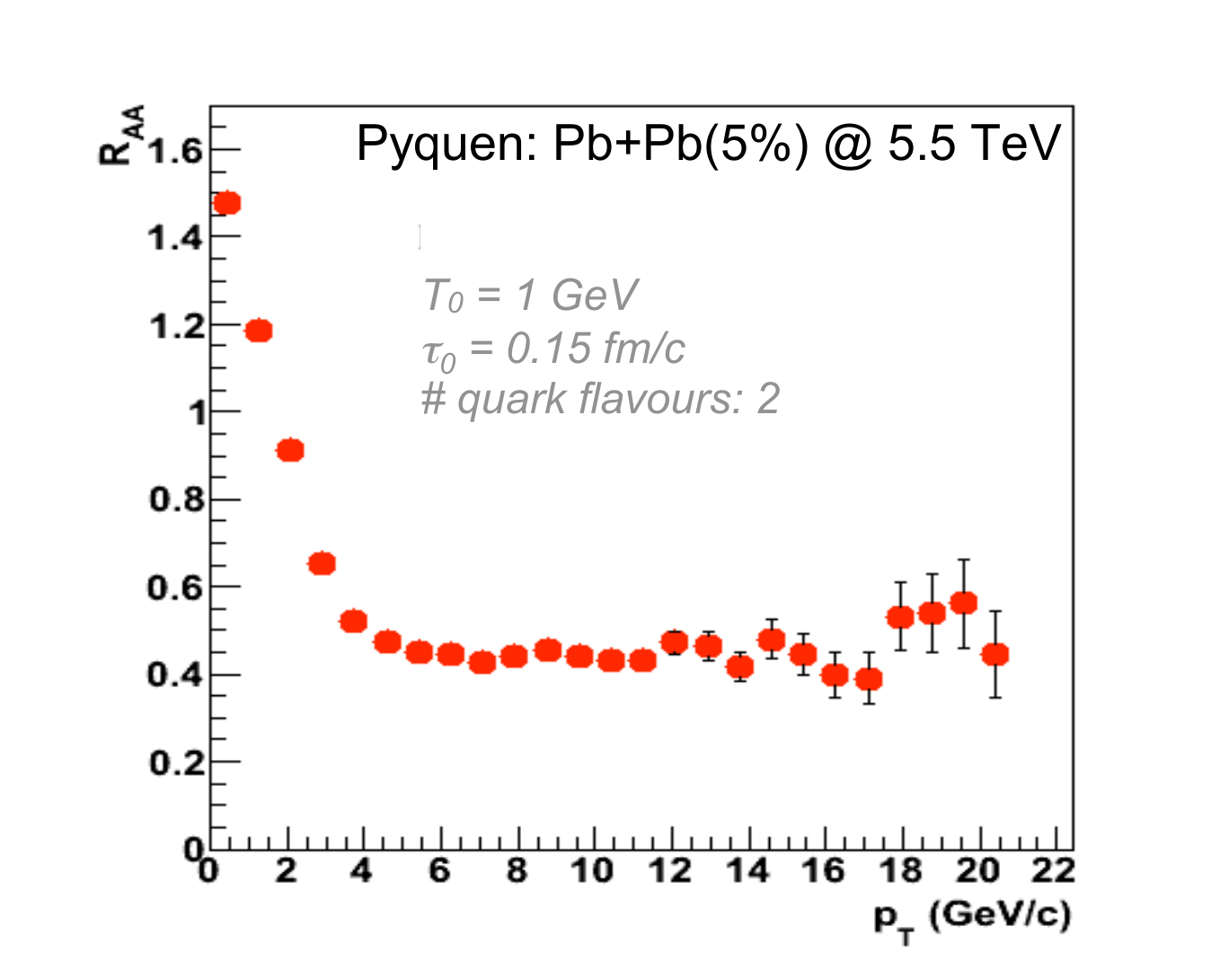}
   \includegraphics[scale=0.5]{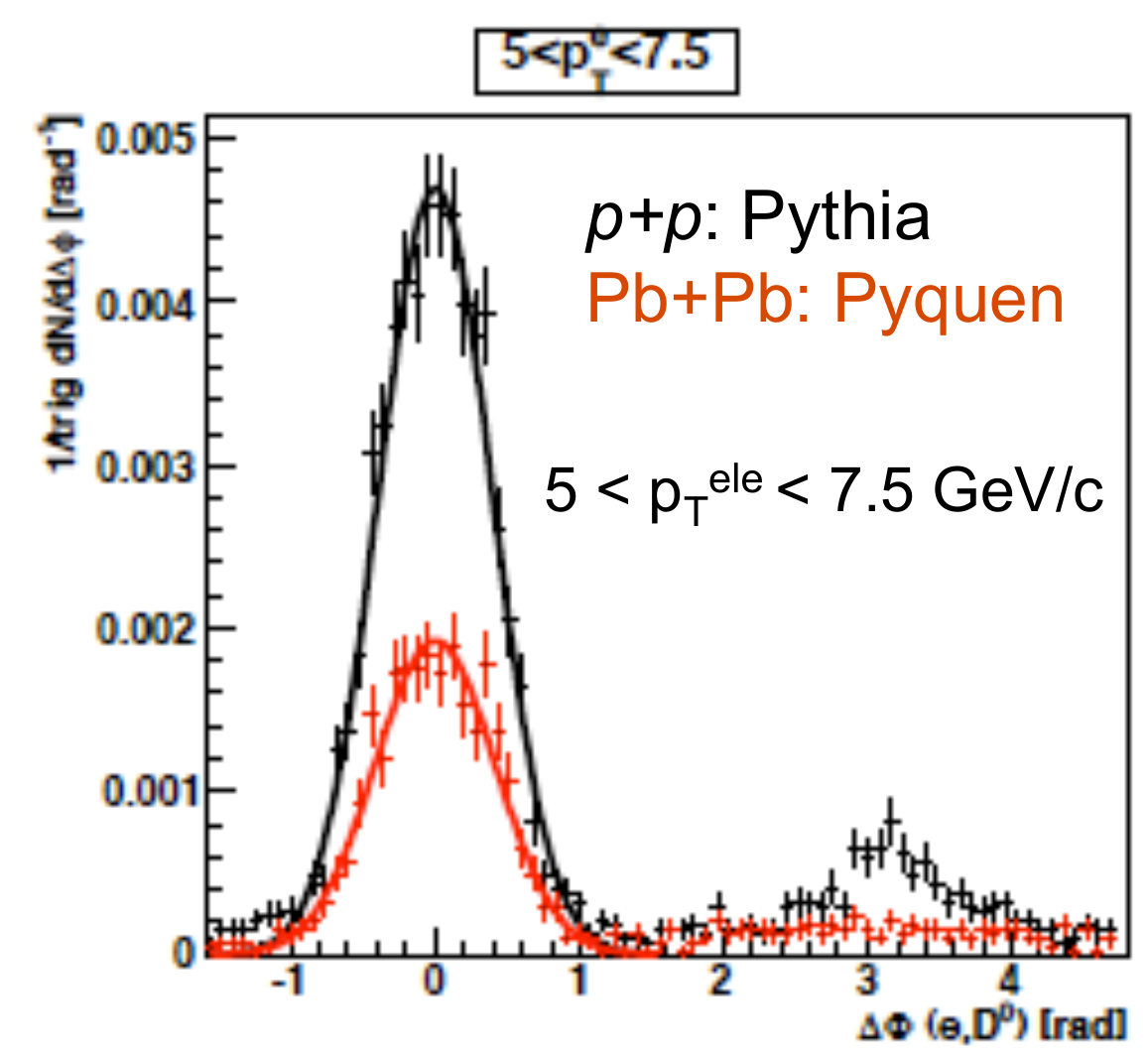}
   \caption{$\raa$ of single electrons (left panel) and $e-D^0$ azimuthal angular correlation distribution (right panel) from Pyquen simulations of central lead-lead collisions at $\snn$~= 5.52 TeV.}  \label{Fig:2}
\end{figure}
ALICE, A Large Ion Collider Experiment, is the dedicated detector to study hot QCD matter in lead-lead at $\snn$~= 5.52 TeV at the CERN-LHC. The initial energy density in the collision zone is expected to be about one order of magnitude higher than at the RHIC facility ($\approx$100~GeV/fm$^3$). The higher energy density allows thermal equilibrium to be reached more quickly and creates a relatively long-lived QGP phase. Therefore, it is expected that most of the in-medium effects will be enhanced. 
Moreover, $\approx$10 times more c${\bar{\rm c}}$ and $\approx$100 times more b${\bar{\rm b}}$ pairs are produced at LHC compared to RHIC energies, providing abundant sensitive probes. 

In ALICE open charmed mesons are fully reconstructed in the hadronic decay channels 
$D^0 \rightarrow K^-\pi^+$ and 
$D^+ \rightarrow K^-\pi^+\pi^+$
in the pseudo-rapidity range $|\eta| <$ 0.9 using their decay topology. 
The decay products of $D$ mesons are detected by the Inner Tracking System and the main tracking device of the ALICE experiment, the Time Projection Chamber. The highly segmented Time-of-Flight detector provides additional charged $K/\pi$ separation in the central detector region.
First results on $D^0$, $D^+$ and $D^+_s$ in $p$+$p$ collisions at $\sqrt{s}$~= 7 TeV were presented in~\cite{alice:carlos}.
$D^{*+}$ mesons are reconstructed through its strong decay into $D^0$ and charged pions.
Figure~\ref{Fig:4}, left panel, shows the $M(K\pi\pi)$-$M(K\pi)$ distribution.
A clear signal is seen at the nominal value of $M(D^{*+})$-$M(D^0)$ with a very good signal-to-background ratio. 

Open heavy-flavour mesons will allow to probe the density of the dense QCD medium via parton energy loss measurements and its predicted quark mass dependence~\cite{tho08}. 
Figure~\ref{Fig:2}, left panel, depicts the nuclear modification factor of single electrons obtained from Pyquen simulations~\cite{igor} of central lead-lead collisions at $\snn$~= 5.52 TeV, which agrees with previous theoretical model calculations~\cite{nestor}.
The $p$+$p$ reference is obtained from Pythia simulations at the same collision energy.
Moreover, it has been shown~\cite{corrMeth:plb09} that jet-like correlations of heavy-flavour particles are sensitive to the production process. First studies using Pyquen simulations are shown in Fig.~\ref{Fig:2}, right panel~\cite{deepame}. The suppression on the away-side indicates the energy loss of heavy quarks in the QCD medium. 

Furthermore, it has been shown that higher order sub-processes like gluon splitting to $c\bar{c}$ pairs may have a significant contribution to the total open charm yield~\cite{gluonrate}. The gluon splitting contribution is studied by identifying the D* content of jets~\cite{alice:dstar}.
By utilizing jet-finder algorithms and isolating the near side of the jet-$D^*$ azimuthal angular correlation, the charm ($D^*$) content in a jet is measured. From the correlation strength on the near side it is possible to estimate the amount of open charm production which results from gluon splitting. The $D^*$ signal from this study is shown in Fig.~\ref{Fig:4}, right panel.

\section{Conclusions}
Heavy quarks are particularly good probes to study the properties of hot QCD matter (especially the transport properties).
RHIC data indicate that the energy loss of heavy quarks in the medium is larger than expected and that in particular bottom is stronger suppressed.
The open charm measurements in proton-proton collisions in ALICE provide an important baseline for the comprehensive studies in lead-lead collisions.
Next-to-leading-order QCD processes, such ass gluon splitting, become important at LHC energies and their contribution can be accesses via the measurements of the {\em charm content in jets}.
Moreover, heavy-flavour particle correlations will allow to study the modification of the fragmentation function in the medium.
First heavy-ion collisions in November 2010 mark the start of the study of hot quark matter in a new energy domain.

\begin{theacknowledgments}
The European Research Council has provided financial support under the European Community's Seventh Framework Programme (FP7/2007-2013) / ERC grant agreement no 210223.
This work was supported in part by a Vidi grant from the Netherlands Organisation for Scientific Research (project number 680-47-232).
\end{theacknowledgments}


\bibliographystyle{aipproc}   

\end{document}